\documentclass{article}
\usepackage{arxiv}

\usepackage[utf8]{inputenc} 
\usepackage[T1]{fontenc}    
\usepackage{hyperref}       
\usepackage{url}            
\usepackage{booktabs}       
\usepackage{amsfonts}       
\usepackage{nicefrac}       
\usepackage{microtype}      
\usepackage{lipsum}		
\usepackage{hyphenat} 
\usepackage{graphicx}
\usepackage[numbers,square]{natbib} 
\usepackage{doi}
\usepackage{longtable}
\usepackage{tabularx}
\usepackage{multirow}
\usepackage{array}
\usepackage{academicons}
\usepackage{xcolor}
\definecolor{orcidgreen}{HTML}{A6CE39}
\newcommand{\orcidlink}[1]{%
\href{https://orcid.org/#1}{%
\raisebox{-0.15ex}{\includegraphics[height=0.8em]{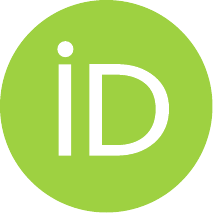}}}}

\title{Clinician Perspectives on Type 1 Diabetes Guidelines and Glucose Data Interpretation}

\author{
\parbox{\textwidth}{
\centering
Mohammed Basheikh\textsuperscript{1}\,\orcidlink{0000-0002-0671-6742}\thanks{CONTACT Mohammed Basheikh. Email: mohammed.basheikh@postgrad.manchester.ac.uk},
Rujiravee Kongdee\textsuperscript{1}\,\orcidlink{0000-0003-0375-2833}\thanks{CONTACT Rujiravee Kongdee. Email: rujiravee.kongdee@manchester.ac.uk}, 
Hood Thabit\textsuperscript{2,3}\,\orcidlink{0000-0001-6076-6997}, \\
Bijan Parsia\textsuperscript{1}\,\orcidlink{0000-0002-3222-7571}, 
Sarah Clinch\textsuperscript{1}\,\orcidlink{0000-0002-9305-4774}, 
Simon Harper\textsuperscript{1}\,\orcidlink{0000-0001-9301-5049} \\[3mm]
\textsuperscript{1}Department of Computer Science, University of Manchester, UK \\
\textsuperscript{2}Diabetes, Endocrine and Metabolism Centre, Manchester University NHS Foundation Trust \\
\textsuperscript{3}Division of Diabetes, Endocrinology and Gastroenterology, University of Manchester, UK \\
}
}

\renewcommand{\headeright}{}
\renewcommand{\undertitle}{}
\renewcommand{\shorttitle}{Clinician Perspectives on Type 1 Diabetes Guidelines and Glucose Data Interpretation}

\usepackage{wasysym}

\hypersetup{
pdftitle={A template for the arxiv style},
pdfsubject={q-bio.NC, q-bio.QM},
pdfauthor={David S.~Hippocampus, Elias D.~Striatum},
pdfkeywords={First keyword, Second keyword, More},
}

\begin{document}
\maketitle
\begin{abstract}

This study explored healthcare professionals’ perspectives on the management of Type 1 Diabetes Mellitus (T1DM) through a two-part questionnaire. The first part examined how clinicians prioritise and apply current clinical guidelines, including the relative importance assigned to different aspects of T1DM management. The second part investigated clinicians’ perceptions of patients’ ability to interpret data from the glucose monitoring devices and to make appropriate treatment decisions. An online questionnaire was completed by 19 healthcare professionals working in diabetes-related roles in the United Kingdom. The findings revealed that blood glucose management is prioritised within clinical guidance and that advice is frequently tailored to individual patient needs. Additionally, clinicians generally perceive that data presented in glucose monitoring devices is easy for patients to interpret and based on these data, they believe that patients occasionally make correct treatment decisions.

\end{abstract}

\keywords{Type 1 Diabetes \and Medical guidelines \and Healthcare professionals \and Self-management \and Glucose monitoring devices \and Interpretation \and Questionnaire}

\section{Introduction}

Type 1 Diabetes Mellitus (T1DM) is a chronic autoimmune disease that attacks the pancreas's beta-cells, inhibiting insulin production and preventing the body from self-regulating blood glucose levels \cite{atkinson2014type}. The Juvenile Diabetes Research Foundation (JDRF) suggest that roughly 400,000 people live with T1DM in the UK \cite{breakthrought1dItsJoke}. It is estimated the National Health Service (NHS) has spent around £1.92 billion in the last decade for diabetes treatment \cite{diabetesTypeDiabetes}. It is considered that due to the chronic complications resulting from poor glycaemic control, the risk of mortality of a person with type 1 diabetes is higher when compared to a non-diabetic counterpart \cite{NICE2015}. For those living with T1DM to achieve the stringent glycaemic control needed to mitigate this increased mortality risk, blood glucose levels need to be monitored regularly and adjusted with either exogenous insulin or consuming a glucose source. Less than 60\% of people with T1DM achieve the desired glycaemic target ranges to mitigate the risk of T1DM complications \cite{sherr2024}.  

The National Institute for Health and Care Excellence (NICE) guideline offers comprehensive care and treatment recommendations for adults with type 1 diabetes \cite{NICE2015}, which many healthcare professionals adopt to clinically manage those with type 1 diabetes. Despite the guideline's comprehensiveness, there are various aspects of diabetes management that healthcare professionals must consider, such as dietary management, physical activity, blood glucose management and mitigating cardiovascular risk.

In addition to guidelines, glucose self-monitoring is crucial for maintaining safe glycaemic levels. Diabetes technologies, which include wearable glucose monitoring devices, have been developed to assist people living with T1DM to reduce glycaemic variation while improving quality of life \cite{oviedo2017review}. Currently, there are three major types of blood glucose monitoring systems: self-monitoring of blood glucose (SMBG), and continuous glucose monitoring (CGM) \cite{cappon2019continuous, haskova2020real}. CGM devices have become widely adopted due to the limitations of SMBG and the convenience of continuous glucose data they provide. SMBG users have to take a drop of blood and place it into a strip of a glucometer to see the reading many times a day, which causes pain and is time-consuming

Despite the convenience of CGM devices, their applications' user interface has been reported to be problematic. Research has shown that the interface demands excessive cognitive effort, is difficult to interpret, and requires substantial time to extract meaningful information effectively \cite{katz_data_2018, katz_questioning_2016, pickup_real-time_2015}. These are concerning because people living with T1DM heavily rely on these user interfaces, and misinterpretation of data may result in delayed or inappropriate treatment, leading to life-threatening situations \cite{erbach_interferences_2016}.

In this study, we sought to understand clinicians' perspectives on implementing recommended guidelines and their perceptions of patients' problems associated with current blood glucose monitoring devices among patients with type 1 diabetes. Accordingly, the focus of this work is two-fold. First, to identify the relative importance that healthcare professionals assign to the advice presented in the NICE guidelines. Second, to explore their opinions regarding problems in understanding the glucose monitoring devices in their patients.


\section{Method}

    \subsection{Participant recruitment}
       To be eligible for the study, participants were required to be healthcare professionals working in a diabetes-related field and familiar with the NICE NG17 (Type 1 diabetes in adults: diagnosis and management) guidelines. Participants were recruited through the Manchester Diabetes Centre. In addition, recruitment notices were posted on official medical websites, including those of the Association of British Clinical Diabetologists and Diabetes UK.

    \subsection{Data collection}
        The study was conducted as an online questionnaire administered via the Qualtrics platform and required approximately 10 minutes to complete. The questionnaire consisted of 3 main parts: (1) Demographics, (2) T1DM guidelines and (3) Glucose monitoring devices. Table \ref{questions-table} presents the list of questions.

        \small
\setlength{\tabcolsep}{4pt}
\renewcommand{\arraystretch}{1.2}

\begin{longtable}{|p{2.2cm}|p{6.2cm}|p{7cm}|}
\hline
\textbf{Part} & \textbf{Question} & \textbf{Question type} \\ \hline
\endfirsthead

\hline
\textbf{Part} & \textbf{Question} & \textbf{Question type} \\ \hline
\endhead

\hline
\multicolumn{3}{r}{\small\emph{Continued on next page}} \\
\endfoot

\endlastfoot

\multirow{2}{*}{\parbox{2.2cm}{Demographic}} &
Please select your position &
Multiple choice (single answer):
\begin{itemize}
    \item[$\ocircle$] Consultant Physician
    \item[$\ocircle$] Specialist Registrar
    \item[$\ocircle$] GP
    \item[$\ocircle$] Practice nurse
    \item[$\ocircle$] Specialist Nurse Practitioner (Diabetes) – Hospital/Community care
    \item[$\ocircle$] Specialist Nurse Practitioner (Diabetes) – Primary/Community care
    \item[$\ocircle$] Dietitian
    \item[$\ocircle$] Other (please specify)
\end{itemize} \\ \cline{2-3}

&
Please specify your workplace or hospital postcode (first three letters; if not in the United Kingdom, please specify your country) &
Free-text response \\ \hline

\multirow{2}{*}{\parbox{2.2cm}{T1DM guidelines}} &
Q1. Do you change your self-management advice for people with type 1 diabetes based on each individual’s needs and condition? &
Multiple choice (multiple answers):
    
    \begin{itemize}
        \item[$\square$] I follow the national standards and guidance.
        \item[$\square$] I follow local practice guidance. 
        \item[$\square$] I always follow the national standards and guidance and rarely change my advice according to patient’s clinical or personal needs.
        \item[$\square$] I always follow local practice guidance and rarely change my advice according to patient’s clinical or personal needs. 
        \item[$\square$] I always adjust my advice according to patient’s clinical or personal needs. 
    \end{itemize}

Please describe further (Free-text response)

\\ \cline{2-3}

&
Q2. Based on current Type 1 Diabetes guidance (e.g., NICE NG17, American Diabetes Association, European Association for the Study of Diabetes), which information do you consider most helpful to support clinical management in adults over 25 years with Type 1 Diabetes? (Rank from 1 = most important to 5 = least important) &
Rank order: 

    \begin{itemize}
        \item Dietary management
        \item Physical activity
        \item Blood glucose management
        \item Insulin therapy and delivery
        \item Hypoglycaemia awareness and management 
    \end{itemize}
    
\\ \hline

\multirow{3}{*}{\parbox{2.2cm}{Glucose monitoring devices}} &
Q3. To what extent do you think blood glucose data presented in continuous and flash glucose monitoring systems are difficult to interpret for patients with type 1 diabetes? &
5-point Likert scale:

    \begin{itemize}
        \item Very difficult to interpret
        \item Difficult to interpret
        \item Neither difficult nor easy to interpret
        \item Easy to interpret
        \item Very easy to interpret
    \end{itemize}

\\ \cline{2-3}

&
Q4. How often do you think patients with type 1 diabetes misinterpret their blood glucose data from monitoring systems? &
5-point Likert scale:

    \begin{itemize}
        \item Never
        \item Rarely
        \item Sometimes
        \item Usually
        \item Always
    \end{itemize}

\\ \cline{2-3}

&
Q5. To what extent do you think patients with type 1 diabetes correctly decide on the action to take in response to their blood glucose levels? &
5-point Likert scale: almost always correct; usually correct; occasionally correct; usually not correct; almost never correct 

    \begin{itemize}
        \item Almost always correct
        \item Usually correct
        \item Occasionally correct
        \item Usually not correct
        \item Almost never correct
    \end{itemize}

\\ \hline
\caption{List of questions and response formats}

\label{questions-table} 
\end{longtable}

    \subsection{Ethical approval}
        This study was reviewed and approved by the University of Manchester Research Ethics Committee before data collection began (Reference number: 2023-15646-27225). All participants provided informed consent for their data to be published.

\section{Findings}

    \subsection{Participants demographics}
        A total of 19 healthcare professionals completed the questionnaire. The number of respondents in each role is outlined in Table \ref{tab:hcp_roles}.
        
        \begin{table}[h]
            \centering
            \begin{tabularx}{\linewidth}{|X|c|}
            \hline
            \textbf{Role} & \textbf{Respondents} \\ \hline
            Consultant Physician & 7 \\ \hline
            Specialist Registrar & 4 \\ \hline
            GP & 2 \\ \hline
            Specialist Nurse Practitioner (Diabetes) – Hospital/Community care & 1 \\ \hline
            Specialist Nurse Practitioner (Diabetes) – Primary/Community care & 2 \\ \hline
            Dietitian & 2 \\ \hline
            Other (Retired GP who worked as medical advisor) & 1 \\ \hline
            \end{tabularx}
            \caption{Number of respondents in each role}
            \label{tab:hcp_roles}
        \end{table}

        In terms of the specified workplace, participants were geographically distributed across several regions in England, with a concentration in the North West (see the Appendix).

    \subsection{T1DM Guidelines}
        Q1: Do you change your self-management advice for people with type 1 diabetes based on each individual’s needs and condition? The results are as follows:
            \begin{itemize}
                \item I always adjust my advice according to patient’s clinical or personal needs: 36.36\% 
                \item I always follow local practice guidance and rarely change my advice according to patient’s clinical or personal needs: 6.06\% 
                \item I always follow the national standards and guidance and rarely change my advice according to patient’s clinical or personal needs: 9.09\% 
                \item I follow local practice guidance: 18.18\% 
                \item I follow the national standards and guidance: 30.30\% 
            \end{itemize}
            
        Q2: Based on current Type 1 Diabetes guidance (eg: National Institute for Health and Care Excellence - NG17, American Diabetes Association or European Association for the Study of Diabetes) which information from these guidance do you consider most helpful to support your clinical management in adults (over age of 25) with Type 1 Diabetes. (Rank the list, '1' is the most important and '5' is the least important) The results are presented in Table \ref{tab:guideline_ranking} :
            \begin{table}[h]
                \centering
                \begin{tabularx}{\linewidth}{X c c c c c}
                \hline
                 & \multicolumn{5}{c}{Rank} \\
                \cline{2-6}
                \textbf{Clinical Management Area} & \textbf{1} & \textbf{2} & \textbf{3} & \textbf{4} & \textbf{5} \\
                \hline
                Dietary management & 0 & 4 & 4 & 10 & 1 \\
                Physical activity & 1 & 2 & 2 & 1 & 13 \\
                Blood glucose management & 9 & 3 & 3 & 3 & 1 \\
                Insulin therapy and delivery & 7 & 8 & 0 & 2 & 2 \\
                Hypoglycaemia awareness and management & 2 & 2 & 10 & 3 & 2 \\
                \hline
                \end{tabularx}
                \caption{Clinician prioritisation of Type 1 diabetes management guidelines}
                \label{tab:guideline_ranking}
            \end{table}

        The results suggest that most clinicians report adapting their self-management recommendations for people with T1DM according to individual clinical and personal circumstances. The questions specifically referred to clinician-led decisions that doctors have the authority to modify. Only a small proportion indicated that they rarely adjust their advice and instead primarily adhere to local or national guidance. These findings suggest that personalised clinical judgement is prioritised in routine practice when clinicians have discretion to individualise care.
        Also, the results show agreement among healthcare professionals regarding the prioritisation of current T1DM guidance. Blood glucose management was considered the most important aspect, while dietary management and physical activity were considered the least important. These findings indicate that dietary management and physical activity are given lower priority in clinical practice.

    \subsection{Glucose Monitoring Devices}
\begin{figure}[!h]
                \centering
                \includegraphics[width=0.7\textwidth]{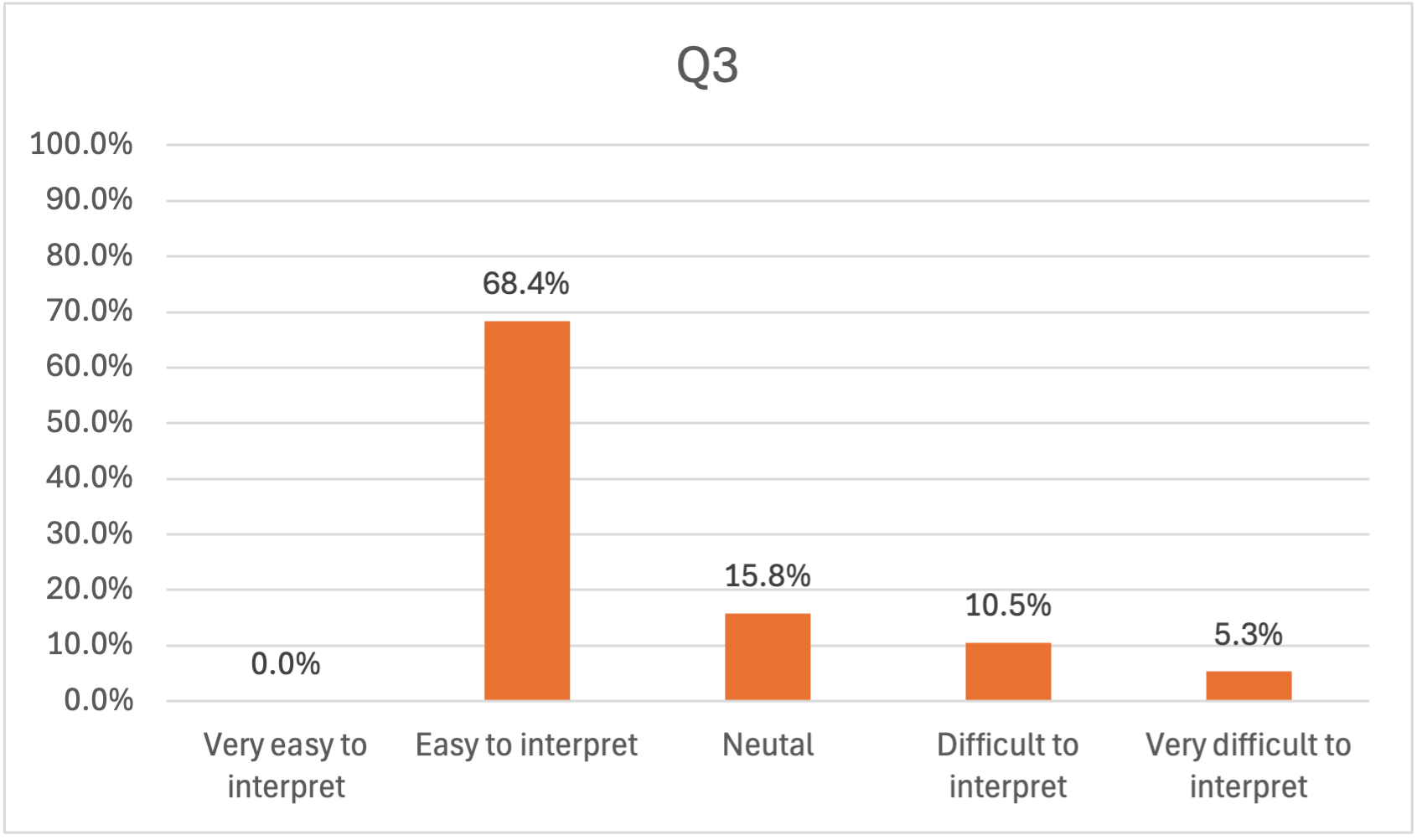}
                \caption{Distribution of respondents’ perceptions of the difficulty of interpreting data presented in continuous and flash glucose monitoring devices (Q3).}
                \label{q3}
            \end{figure}

        Q3: To what extent do you think blood glucose data presented in continuous and flash glucose monitoring systems are difficult to interpret for patients with type 1 diabetes? The results are depicted in Figure \ref{q3}.

\begin{figure}[!h]
                \centering
                \includegraphics[width=0.7\textwidth]{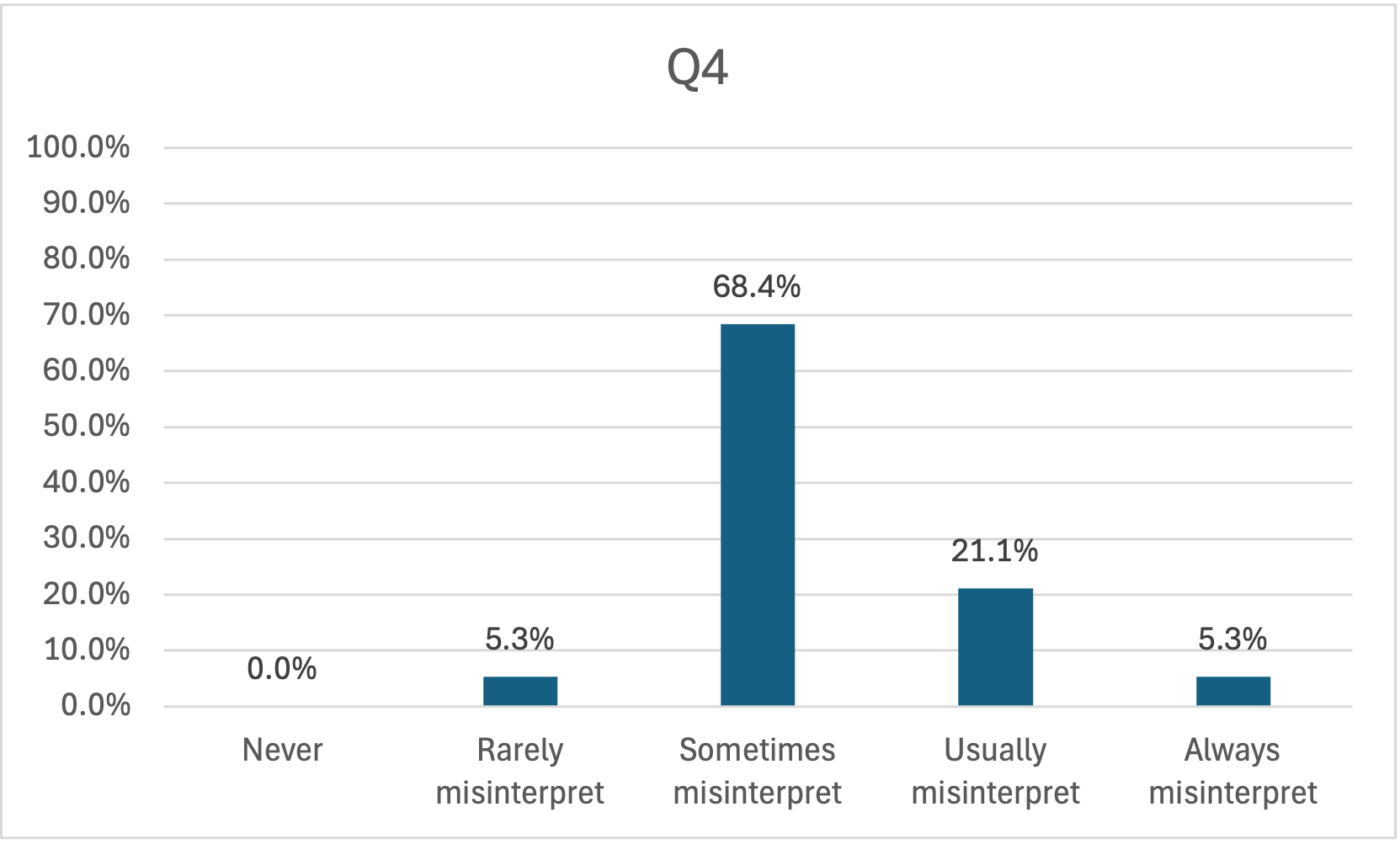}
                \caption{Distribution of respondents' perceptions of the frequency that patients misinterpret their data presented in monitoring devices (Q4).}
                \label{q4}
            \end{figure}

        Q4: How often do you think patients with type 1 diabetes misinterpret their blood glucose data from monitoring systems? The results are shown in Figure \ref{q4}.

\begin{figure}[!h]
                \centering
                \includegraphics[width=0.7\textwidth]{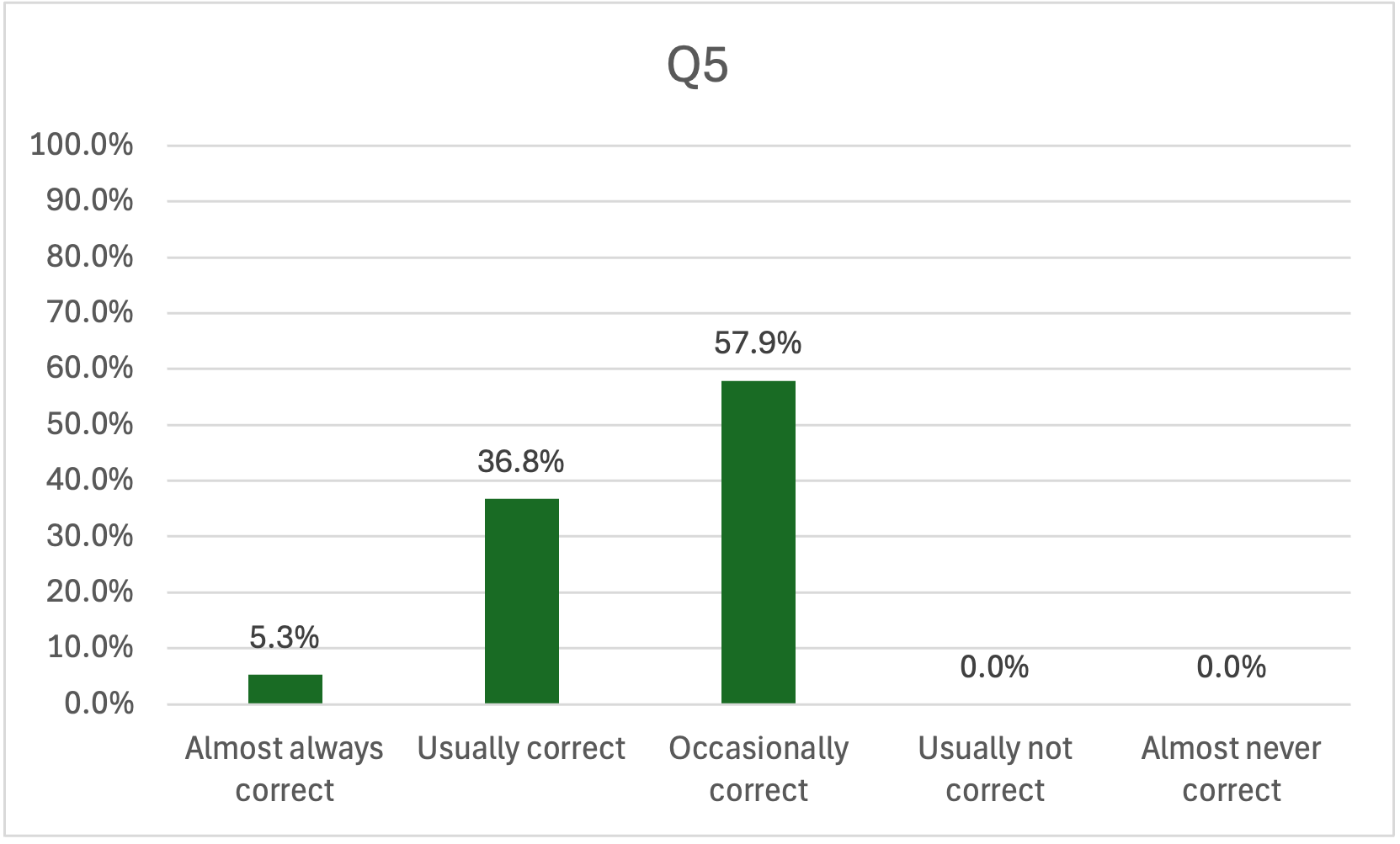}
                \caption{Distribution of respondents' perceptions of the frequency that patients correctly make treatment decisions based on their glucose levels (Q5).}
                \label{q5}
            \end{figure}

        Q5: To what extent do you think patients with type 1 diabetes correctly decide on the action to take in response to their blood glucose levels? The results are presented in Figure \ref{q5}.

        The results suggest that the majority of respondents believe that interpreting data presented by CGM and/or Flash is easy for patients. However, they also consider that patients sometimes misinterpret the data. With respect to treatment decision-making, most respondents think that patients only occasionally make correct decisions.

        The belief that data presented by CGM and/or Flash systems should be easy for people with T1DM to interpret contrasts with findings from empirical research. In one study, semi-structured interviews were conducted with participants with T1DM, who were asked to provide spontaneous verbal interpretations of CGM and Flash user interfaces. The results showed that participants correctly interpreted the displayed glucose data in fewer than 40\% of cases. Furthermore, the correctness of treatment decisions was even lower, at 22\% \cite{kongdee2025glucose}. These findings suggest that, although healthcare professionals may perceive such data as straightforward to interpret, people with T1DM often experience substantial difficulty. Consistent with this, prior research has also reported that while these individuals face challenges in interpreting these devices, healthcare providers tend to assume that this task is straightforward \cite{raj2017understanding}.
        
        This highlights a clear misalignment between patient and clinician perspectives. There is a need for healthcare professionals to reassess their assumptions, as patients may not interpret glucose data as effectively as clinicians expect. Moreover, other factors such as patients' age, native language and socioeconomic background could also affect the interpretability of their device data.
        These insights can be used to support more targeted and effective clinical conversations to help patients understand their data more accurately.

\section{Data Availability}
The dataset is available in a Zenodo Repository `Clinicans-perspective-T1DM' at https://doi.org/10.5281/zenodo.18404347

\section{Acknowledgement}
The authors would like to thank all participants for their time and contribution to this study. Mohammed Basheikh was supported by the University of Jeddah, Saudi Arabia. Rujiravee Kongdee was supported by the Royal Thai Government Scholarship.

\bibliographystyle{acm}
\bibliography{references}  

\appendix
\section{Appendix}

The number of respondents in each location was as follows:
\begin{itemize}
    \item \textbf{Manchester}: 6 
    \item \textbf{London}: 2 
    \item \textbf{Maidstone, Kent}: 1
    \item \textbf{Rishton, Lancashire}: 1
    \item \textbf{Liverpool}: 1 
    \item \textbf{Peterborough}: 1 
    \item \textbf{Sheffield}: 1 
    \item \textbf{Redhill}: 1 
    \item \textbf{Worthing}: 1 
    \item \textbf{Stevenage}: 1 
    \item \textbf{Welwyn Garden City}: 1 
    \item \textbf{Birmingham}: 1
    \item \textbf{St Helens}: 1 
    \item \textbf{Southend-on-Sea}: 1 
\end{itemize}

\end{document}